\documentclass{article}
\usepackage{graphicx} 
\usepackage{subcaption} 
\usepackage{amsmath,amssymb,amsfonts} 
\usepackage{bbm,bm}
\usepackage{xcolor}
\usepackage{comment}
\usepackage{algorithm}
\usepackage{algorithmic}
\usepackage{url,geometry}

\newtheorem{lemma}{Lemma}

\newtheorem{proposition}{Proposition}

\usepackage{setspace}
\title{Constrained Bayesian Optimization under Bivariate Gaussian Process with Application to Cure Process Optimization}

\author{Yezhuo Li and Qiong Zhang\\School of Mathematical and Statistical Sciences, Clemson University
\\Madhura Limaye\\ Manufacturing Science Division, Oak Ridge National Laboratory
\\Gang Li\\Department of Mechanical Engineering, Clemson University}
\date{}

\begin{document}
\maketitle

\begin{abstract}

{\small Bayesian Optimization, leveraging Gaussian process models, has proven to be a powerful tool for minimizing expensive-to-evaluate objective functions by efficiently exploring the search space. Extensions such as constrained Bayesian Optimization have further enhanced Bayesian Optimization’s utility in practical scenarios by focusing the search within feasible regions defined by a black-box constraint function. 
However, constrained Bayesian Optimization in is developed 
based on the independence Gaussian processes assumption between objective and constraint functions, which may not hold in real-world applications.
To address this issue, we use the bivariate Gaussian process model to characterize the dependence between the objective and constraint functions and developed the constrained 
expected improvement acquisition function under this model assumption. 
We show case the performance of the proposed approach with an application to cure process optimization in Manufacturing.  }
\end{abstract}

\section*{Keywords}
Bayesian Optimization, Constrained Bayesian Optimization, Gaussian Process, Bivariate Distribution.

\section{Introduction}
The development of manufacturing processes often relies on 
exploring the parameter settings of simulation models 
to find the optimal decision 
with respect to certain objectives. For example, based on the simulation experiments of laminated composite cure processes, optimizing parameters such as temperature, pressure, and cure time is crucial for ensuring product quality and minimizing defects, as demonstrated in previous studies on composite manufacturing \cite{gutowski1994development}. 
Due to the complex structure of simulations of manufacturing processes, the investigation of mathematical properties of the resulting objective functions often encounter unique challenges.
Therefore, the objective functions are often treated as black-box functions in practice \cite{kirkpatrick1983optimization} and derivative-free optimization approaches are often used for decision-making with manufacturing process simulations.


Bayesian Optimization (e.g., \cite{Frazier2018}) has emerged as a promising approach for optimizing black-box objective functions. 
Bayesian Optimization constructs a surrogate model of the objective function, typically using Gaussian Process regression, and selects parameter values for simulation evaluations through an acquisition function that balances exploration and exploitation \cite{brochu2010tutorial, snoek2012practical}. This framework has proven effective in high-cost evaluation settings by efficiently using the limited number of computational budget to achieve high quality solutions \cite{jones1998efficient, mockus1994bayesian}. Gardner et al. \cite{gardner2014bayesian} further 
extended the BO framework for constrained
Bayesian Optimization with black-box inequality constraints. constrained
Bayesian Optimization incorporates a feasibility indicator to guide the search, prioritizing feasible regions and reducing evaluations in infeasible areas. This approach has been successfully applied to some challenging tasks, such as hyperparameter tuning in machine learning, where feasible regions are often small and costly to explore \cite{snoek2015scalable}. However, constrained
Bayesian Optimization in \cite{gardner2014bayesian} assumes independence between the Gaussian process surrogates of objective and constraint functions, which may not hold real-world applications.

We include a bivariate Gaussian process model under the framework of constrained Bayesian Optimization in \cite{gardner2014bayesian}. This model is able to characterize linear correlation between the objective and the constraint functions through a separable covariance function
\cite{branke2008multiobjective, alvarez2011computationally}. Under the proposed model, we derived the expected constrained improvement based
on the definition in \cite{gardner2014bayesian}.
The proposed approach is motivated by an application in composite curing, where interdependent factors—time and temperature—must be carefully managed to minimize deformation. In this context, the objective function represents the deformation of the composite material, while the inputs are the process parameters: time and temperature. Optimizing this process is challenging due to the complex relationship between these parameters and deformation behavior.
We compare the proposed method with the original constrained Bayesian Optimization approach using the application of the cure process optimization. The results show that, the proposed approach performs similar to the constrained Bayesian Optimization algorithm without characterizing the correlation between the objective and the constraint functions. This finding suggests that the separable covariance function may not be accurate for this application example, and we should investigate further improvement to the covariance function. 

\section{Problem Description}
Consider the problem of minimizing function  $y(\bm x)$ subject to a constraint $z(\bm x)$ with an input vector $\bm x=(x_1, \ldots, x_d)^\top\in\mathcal X\subset \mathbb R^d$:
\begin{equation}\label{eq:optimization}
    \min \quad y(\bm x) \quad
    \mathrm{s.t.} \quad\ z(\bm x) \geq c  \quad \bm x \in \mathcal{\bm X}
\end{equation}
where threshold value $c$ in the constraint is known. 
For simulation-based optimization problems in manufacturing, 
both the objective and constraint functions are often black-box functions with unknown mathematical properties to utilize gradient based optimization approaches. 
Also, the number of function evaluations is constrained by a limited computational budget, prohibiting exhaustive search methods. 
Bayesian Optimization (BO, \cite{jones1998efficient, Frazier2018}) is a popular approach to tackle this type of optimization problems. 
In Algorithm \ref{alg:optimization}, we summarize the implementation
of Bayesian Optimization for a constraint problem in \eqref{eq:optimization}. The two key components in this algorithm is the surrogate model and the aquisition function.

\begin{algorithm}[!h]
\caption{An Algorithm for Constraint Bayesian Optimization with a blackbox Constraint}
\label{alg:optimization}
\begin{algorithmic}[1]
\STATE \textbf{Preliminary:} Statistical surrogate models for the objective function $y(\bm x)$ and the constraint function $z(\bm x)$
\FOR{n= 1 \textbf{to} $N$  }
    \STATE Compute acquisition function based on the model assumptions of $y(\bm x)$ and $z(\bm x)$, and find $\bm x_{n}$ that maximizes the acquisition over $\bm x \in \mathcal{X}$.
    \STATE Evaluate $y(\bm x)$ and $z(\bm x)$ at $\bm x_{n}$ and update the surrogate models  of $y(\bm x)$ and $z(\bm x)$ based on the new evaluations.
\ENDFOR
\STATE \textbf{Return} Provide optimal solution based on the  surrogate models of $y(\bm x)$ and $z(\bm x)$.
\end{algorithmic}
\end{algorithm}

For Bayesian Optimization without a blackbox constraint, Gaussian process (e.g., \cite{sacks1989designs}) is a popular choice of surrogate models and expected improvement \cite{jones1998efficient} is often used as the acquisition function. 
Under this framework, the response of the blackbox function $y(\bm x)$  is modeled as a realization of the stochastic process:
\begin{equation}\label{eq:gp}
y(\bm x) = \mu + \epsilon(\bm x)
\end{equation}
where  $\mu$  represents the unknown deterministic mean, and  $\epsilon(\bm x)$ , defined over  $\bm x \in \mathcal{X}$ , is a zero-mean Gaussian Process with variance  $\sigma^2$  and correlation function  $R(\cdot, \cdot)$ . 
We assume that the function  $y(\bm x)$  has been evaluated at $n$ input points  $\bm x_1, \bm x_2, \dots, \bm x_n$, resulting in the corresponding outputs  $\mathbf{y}_n = (y(\bm x_1), y(\bm x_2), \dots, y(\bm x_n))^\top$. Under the Gaussian Process framework, the conditional distribution of  $y(\bm x)$ at a new input $\bm x$, given the observed data  $\mathbf{y}_n$, can be represented as follows \cite{ jones1998efficient, schon2011manipulating}:
\begin{equation}\label{eq:cond}
y(\bm x) \mid \mathbf{y}_n \sim \mathcal{N}(\hat{y}(\bm x), s^2(\bm x)),
\end{equation}
with mean and variance 
\begin{equation}\label{eq:mean}
\hat{y}(\bm x) = \hat{\mu} + \bm r(\bm x)^\top \bm R^{-1} (\mathbf{y}_n - \mathbf{1}\hat{\mu})\quad \mathrm{and} \quad
s^2(\bm x) = \sigma^2 \left( 1 - \bm r(\bm x)^\top \bm R^{-1} \bm r(\bm x) - \frac{\big(1 - \mathbf{1}^\top \bm R^{-1} \bm r(\bm x)\big)^2}{\mathbf{1}^\top \bm R^{-1} \mathbf{1}} \right),
\end{equation}
where  $\hat{\mu} = \mathbf{1}^\top \bm R^{-1} \mathbf{y}_n/\mathbf{1}^\top  \bm R^{-1} \mathbf{1}
$, $\bm r(\bm x)= (R(\bm x, \bm x_1), \cdots, R(\bm x, \bm x_n))^\top$, $\bm R$ denotes the $n \times n$ matrix where the $(i, j)$-th entry is $R(\bm x_i, \bm x_j)$ representing the correlation between $\epsilon(\bm x_i)$ and $\epsilon(\bm x_j)$, and $\mathbf{1}$ represents an $n$-dimensional vector of ones. 
The acquisition function, expected improvement \cite{jones1998efficient} is constructed by
    \begin{align}
        \mathrm{EI}(\bm x) = \mathbb{E}
        \left\{\max \{ 0, y_{\min} - y(\bm x) \}  \right\},\label{eq:EI}
    \end{align}
where $y_{\min}$ represents the current minimum observation and
the expectation is taken with respect to the conditional distribution of $y(\bm x)$ in \eqref{eq:cond}.

For the problem in \eqref{eq:optimization}, \cite{gardner2014bayesian} assumes that $y(\bm x)$ and $z(\bm x)$ are realizations of two independent Gaussian processes:
\begin{equation}\label{eq:bivariate}
    y(\bm x) = \mu_y + \epsilon_y(\bm x), 
    \quad z(\bm x) = \mu_z + \epsilon_z(\bm x)
\end{equation}
where $\mu_y$ and $\mu_z$ are constants, and $\epsilon_y(\bm x)$ and $\epsilon_z(\bm x)$ are modeled as a zero-mean independent Gaussian Processes with variances $\sigma^2_y$ and $\sigma^2_z$, respectively. 
By evaluating $y(\bm x)$ and $z(\bm x)$ at $\bm x_1, \ldots, \bm x_n$, 
we collected  $\mathbf{y}_n =\begin{bmatrix}
    y(\bm x_1), \cdots, y(\bm x_n)
\end{bmatrix}^\top$ and $\mathbf{z}_n=\begin{bmatrix}
    z(\bm x_1), \cdots, z(\bm x_n)
\end{bmatrix}^\top$. Therefore, we can obtain conditional distributions 
$y(\bm x)|\bm y_n$ and $z(\bm x)|\bm z_n$ similarly as in \eqref{eq:cond}. The acquisition function of expected contained improvement \cite{gardner2014bayesian} is defined by 
    \begin{equation}
        \mathrm{ECI} (\bm x) = \mathbb{E}
        \left\{\max \{ 0, y_{\min} - y(\bm x) \}  I \left( z(\bm x) \ge c \right) \right\}=\mathbb{E}
        \left\{\max \{ 0, y_{\min} - y(\bm x) \}\right\} \mathbb{P}\left\{ \left(z(\bm x) \ge c \right) \right\}\label{eq:EI_c},
    \end{equation}
    where the expectation and probability are computed based on the conditional distributions of $y(\bm x)|\bm y_n$ and $z(\bm x)|\bm z_n$, respectively.
In this paper, we model the dependence between $y(\bm x)$
and $z(\bm x)$, and construct the acquisition function of expected contained improvement based on the joint conditional distribution of 
$y(\bm x)$
and $z(\bm x)$. We describe the proposed surrogate model and the acquisition function in sections \ref{sec:parameters} and \ref{sec:EC}, respectively.

\section{Bivariate Gaussian Process Models}\label{sec:parameters}
Consider that a pair of $y(\cdot)$ and $z(\cdot)$ is a realization of the bivariate Gaussian process. Following \eqref{eq:bivariate}, we modify the independent assumption of $\varepsilon_y(\bm x)$ and $\varepsilon_z(\bm x)$, and assume that 
\begin{equation}
\text{Cov}(y(\bm x), y(\bm x^{\prime})) = \sigma_y^2 R(\bm x, \bm x^{\prime}),  \quad \text{Cov}(z(\bm x), z(\bm x^{\prime})) = \sigma_z^2 R(\bm x, \bm x^{\prime}), \quad \text{Cov}(y(\bm x), z(\bm x^{\prime})) = \rho \sigma_y \sigma_z R(\bm x, \bm x^{\prime}),\label{eq:3covs}
\end{equation}
where $R(\cdot, \cdot)$ is the correlation function defined similarly in \eqref{eq:gp}. 
Under the bivariate Gaussian process assumption, we have that
    \begin{align}
        y(\bm x) \sim \mathcal{N}(\mu_y, \sigma_y^2), \quad z(\bm x) \sim \mathcal{N}(\mu_z, \sigma_z^2), \quad \begin{bmatrix}
y(\bm x) \\
z(\bm x)
\end{bmatrix}
 \sim \mathcal{MVN}\left( \begin{bmatrix}
\mu_y \\
\mu_z
\end{bmatrix}, \begin{bmatrix}
\sigma_y^2 & \rho \sigma_y \sigma_z \\
\rho \sigma_z \sigma_y & \sigma_z^2
\end{bmatrix} \right), \label{eq:3distributions}
\end{align}

Under the bivariate Gaussian process assumption, 
the responses $\mathbf{y}_n =\begin{bmatrix}
    y(\bm x_1), \cdots, y(\bm x_n)
\end{bmatrix}^\top$ and $\mathbf{z}_n=\begin{bmatrix}
    z(\bm x_1), \cdots, z(\bm x_n)
\end{bmatrix}^\top$ follow  a multivariate normal distribution such that 
\begin{align}
        \begin{bmatrix}
\mathbf{y}_n \\
\mathbf{z}_n
\end{bmatrix}
 \sim \mathcal{MVN} \left( \begin{bmatrix}
    \mu_y \mathbf{1} \\
    \mu_z \mathbf{1}
\end{bmatrix}, \begin{bmatrix}
\sigma_y^2 & \rho \sigma_y \sigma_z \\
\rho \sigma_z \sigma_y & \sigma_z^2
\end{bmatrix} \otimes \bm R\right),
\label{eq:gp2}
\end{align}
 where $\bm R$ is the correlation matrix in as in \eqref{eq:mean}, and $\otimes$ is Kronecker product. The correlation parameter $\rho$ characterizes the dependence between $y(\bm x)$ and $z(\bm x)$.

Given $\mu_y$, $\mu_z$, $\sigma_y^2$ and $\sigma_z^2$, the conditional distributions can be expressed by
\begin{equation}
y(\bm x) \mid \mathbf{y}_n, \mathbf{z}_n \sim \mathcal{N} \left( \mu_{y\mid \mathbf{y}_n, \mathbf{z}_n}, \Sigma_{y\mid \mathbf{y}_n, \mathbf{z}_n} \right) \quad\mathrm{and}\quad
\quad z(\bm x) \mid \mathbf{y}_n, \mathbf{z}_n \sim \mathcal{N} \left( \mu_{z\mid \mathbf{y}_n, \mathbf{z}_n}, \Sigma_{z\mid \mathbf{y}_n, \mathbf{z}_n} \right)
    \label{eq:conditional distribution_y}
\end{equation}
where \[
\mu_{y\mid \mathbf{y}_n, \mathbf{z}_n}=\mu_y+\widetilde{\bm r}(\bm x)_y^\top \widetilde{\bm R}^{-1} \begin{bmatrix}
    \mathbf{y}_n-\mu_y \bm 1\\
    \mathbf{z}_n-\mu_z \bm 1
\end{bmatrix},\quad
\mu_{z\mid \mathbf{y}_n, \mathbf{z}_n}=\mu_z+\widetilde{\bm r}(\bm x)_z^\top \widetilde{\bm R}^{-1} \begin{bmatrix}
    \mathbf{y}_n-\mu_y \bm 1\\
    \mathbf{z}_n-\mu_z \bm 1
\end{bmatrix},
\]
and
\[
\Sigma_{y\mid \mathbf{y}_n, \mathbf{z}_n}= \sigma_y^2-\widetilde{\bm r}(\bm x)_y^\top \widetilde{\bm R}^{-1} \widetilde{\bm r}(\bm x)_y + \frac{\sigma_y^2 \left( 1-\bm 1^\top \bm R^{-1} \bm r(\bm x) \right)^2}{\bm 1^\top \bm R^{-1} \bm 1},
\quad
\Sigma_{z\mid \mathbf{y}_n, \mathbf{z}_n}=\sigma_z^2-\widetilde{\bm r}(\bm x)_z^\top \widetilde{\bm R}^{-1} \widetilde{\bm r}(\bm x)_z + \frac{\sigma_z^2 \left( 1-\bm 1^\top \bm R^{-1} \bm r(\bm x) \right)^2}{\bm 1^\top \bm R^{-1} \bm 1},
\]
with 
\[
\widetilde{\bm r}(\bm x)_y= \left[\begin{bmatrix}
    \sigma^2_y & \rho 
    \sigma_y \sigma_z
\end{bmatrix} \otimes \bm r (\bm x) \right]^\top,\quad\widetilde{\bm r}(\bm x)_z= \left[\begin{bmatrix}
    \rho 
    \sigma_y \sigma_z & \sigma^2_z
\end{bmatrix}\otimes \bm r (\bm x) \right]^\top\quad\mathrm{and} \quad  \widetilde{\bm R}=\begin{bmatrix}
\sigma_y^2 & \rho \sigma_y \sigma_z \\
\rho \sigma_z \sigma_y & \sigma_z^2
\end{bmatrix} \otimes \bm R.\]

By maximizing the log likelihood function from \eqref{eq:3distributions}, the estimated means can be expressed as
\begin{equation}
    \hat{\mu}_y = \left( \mathbf{1}^\top (\sigma_y^{2} \bm R)^{-1} \mathbf{1} \right)^{-1} \mathbf{1}^\top (\sigma_y^{2} \bm R)^{-1} \mathbf{y}_n, 
    \quad  \hat{\mu}_z = \left( \mathbf{1}^\top (\sigma_z^{2} \bm R)^{-1} \mathbf{1} \right)^{-1} \mathbf{1}^\top (\sigma_z^{2} \bm R)^{-1} \mathbf{z}_n
\end{equation}

Similarly, by maximizing log likelihood function from \eqref{eq:3distributions}, the estimated variances are
\begin{equation}
    \hat{\sigma}_y^{2} = \frac{(\mathbf{y}_n -\hat{\mu}_y \mathbf{1})^\top ( \bm R)^{-1} (\mathbf{y}_n-\hat{\mu}_y \mathbf{1})}{n}, \label{eq:sigma_Y}
    \quad  \hat{\sigma}_z^{2} = \frac{(\mathbf{z}_n-\hat{\mu}_z \mathbf{1})^\top ( \bm R)^{-1} (\mathbf{z}_n-\hat{\mu}_z \mathbf{1})}{n}
\end{equation}
Also, the estimated $\rho$ can be derived by maximizing the likelihood function in \eqref{eq:gp}:
\begin{equation}
    \hat{\rho} = \arg \max \log L(\rho \mid \mathbf{y}_n, \mathbf{z}_n, \hat{\mu}_y, \hat{\mu}_z, \hat{\sigma}_y^{2}, \hat{\sigma}_z^{2}, \bm R).
\end{equation}

The correlation coefficient $\rho$ between $y(\mathbf{x})$ and $z(\mathbf{x})$ is essential in the bivariate Gaussian process model as it captures the dependency between the two functions. $\rho$ explicitly defines the off-diagonal terms in the covariance matrix, ensuring the model accounts for inter-output correlation. 

\section{Acquisition Function for Bivariate Gaussian Process Models} \label{sec:EC}
In this section, we introduce an acquisition function designed to sequentially select the next experimental setting. Specifically, we derive a closed-form expression of the expected constrained improvement (ECI) \eqref{eq:EI_c} under the Bivariate Gaussian process assumption described in Section \ref{sec:parameters}.
By incorporating the correlation between $y(\bm x)$ and $z(\bm x)$, the expression of ECI in \eqref{eq:EI_c} can not be separated into the expectations with respect to $y(\bm x)$ and $z(\bm x)$, respectively. 
We derive the expression of ECI under the bivariate Gaussian process in the following proposition.

\begin{proposition} \label{prop1} Under bivariate Gaussian process model in Section \ref{sec:parameters} and the property in Lemma \ref{lemma:bivariate_normal}, the closed-form expression of $\mathrm{ECI}(\bm x)$ in \eqref{eq:EI_c} is given by
\begin{equation*}
        \mathrm{ECI}(\bm x) = (t_1(\bm x) + t_2(\bm x)) \times t_3(\bm x)
\end{equation*}
where 
\[
        t_1(\bm x) = y_{\min} - \mu_y-\rho \sigma_y \frac{\phi\left( \frac{c-\mu_z}{\sigma_z} \right)}{1-\Phi\left( \frac{c-\mu_z}{\sigma_z} \right)},\quad
t_2(\bm x) = \sigma_y \sqrt{1-\rho^2} \frac{\phi\left( \frac{t_1(\bm x)}{\sigma_y \sqrt{1-\rho^2}} \right)}{\Phi\left( \frac{t_1(\bm x)}{\sigma_y \sqrt{1-\rho^2}} \right)},\]
and\[
t_3(\bm x) = \Phi\left( \frac{y_{\min}-\mu_y}{\sigma_y} \right) - \Phi_2\left( \frac{y_{\min}-\mu_y}{\sigma_y}, \frac{c-\mu_z}{\sigma_z}; \rho\right),\]
with $\phi \left( \cdot \right)$ and $\Phi \left( \cdot \right)$   be the probability density function (PDF) and the cumulative density function (CDF) of the standard normal distribution and $\Phi_2(\cdot, \cdot ; \cdot)$  refers to the bivariate cumulative distribution function of a standard bivariate normal distribution with a correlation coefficient $\rho$ between the two random variables. 
\end{proposition}
The proof of this proposition is deferred to the appendix. 
When the CDF values $\Phi(\cdot)$ approaching 1 or 0, there are potential numerical issues. The reason is that the expected improvement becomes nearly zero, and further evaluations may result in numerical instability. We provide approximation to address this issue, and our implementation of the algorithm is based on the following approximations:
As $\Phi((c-\mu_z)/\sigma_z)\rightarrow 1$,  
$
\mathrm{ECI} (\bm x) \rightarrow (t_1^\prime (\bm x) + t_2^\prime (\bm x)) \times t_3(\bm x),
$
where 
\begin{equation*}
    t_1^\prime(\bm x) = y_{\min} - \mu_y-\frac{\rho \sigma_y (c-\mu_z)}{\sigma_z}
    \quad\mathrm{and}\quad t_2^\prime(\bm x) = \sigma_y \sqrt{1-\rho^2} \frac{\phi\left( \frac{t_1^\prime (\bm x)}{\sigma_y \sqrt{1-\rho^2}} \right)}{\Phi\left( \frac{t_1^\prime (\bm x)}{\sigma_y \sqrt{1-\rho^2}} \right)} 
\end{equation*}
Also, as $\Phi\left( t_1(\bm x)/(\sigma_y \sqrt{1-\rho^2}) \right) \to 0$, $\mathrm{ECI} (\bm x) \rightarrow 0$.

\section{Application to Cure Process Optimization}

Cure processes are popularly used in the manufacturing of
thermoset based fiber reinforced composite laminates. In the simulation of cure process, it is essential to optimize the parameter setting of the cure process to reduce the deformation in the final product \cite{limaye2024computational}. 
We consider a case study of the cure of the L-shaped laminate in \cite{paper2}. The optimization problem is given by \eqref{eq:optimization},
where the decision variable $\bm x=(T_1, t_1, T_2, t_2)^\top$ are the two change points in the cure cycle, the objective function $y(\bm x)$ is the deformation in the final product, and the constraint function $z(\bm x)$
is the degree of cure, with the constant $c$ being 0.96. 

\begin{figure}[ht]
  \begin{subfigure}{.5\textwidth}
  \centering
    \includegraphics[width=0.9\linewidth]{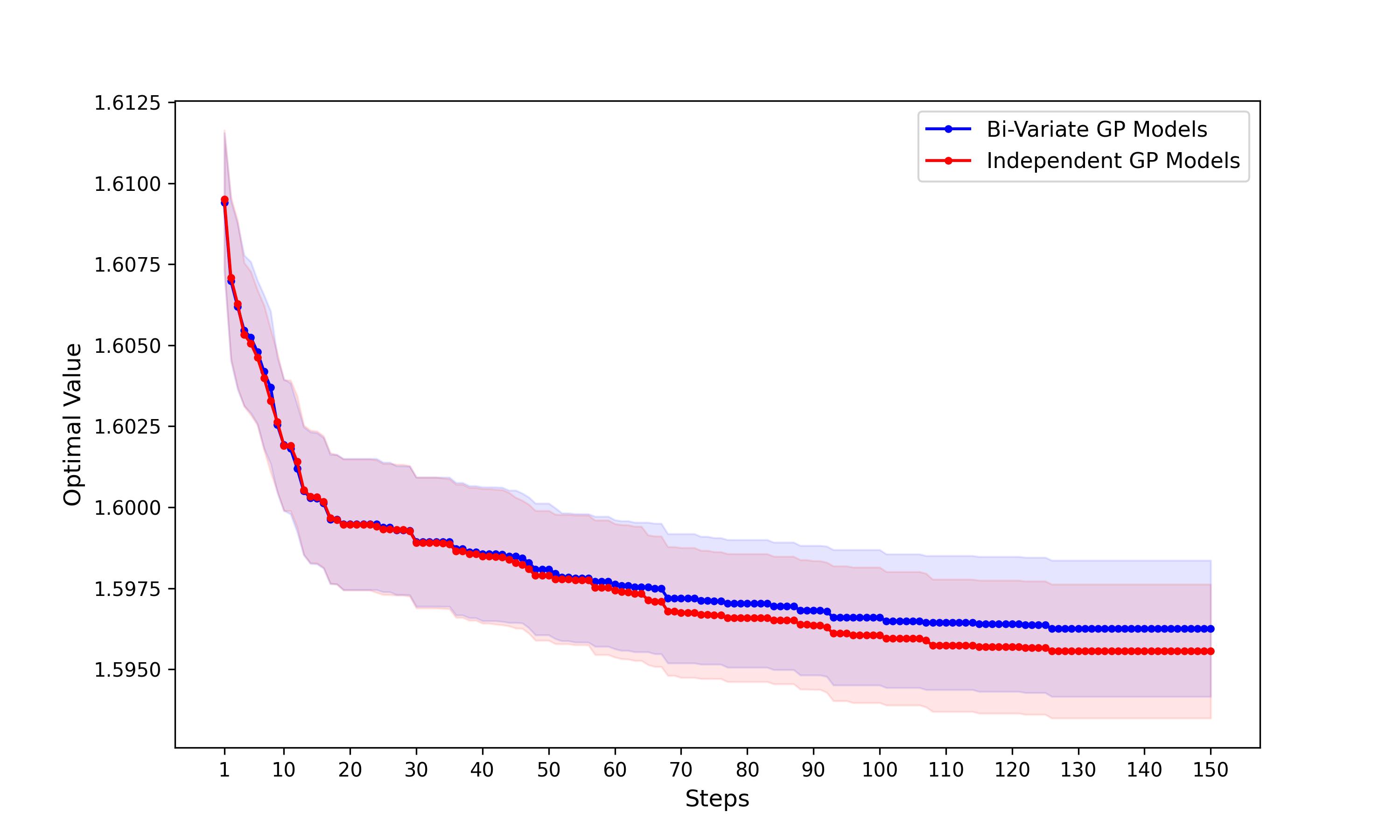}
    \caption{Initial Sample Size $25$.}
    \label{fig:25_initial_sample_size}
  \end{subfigure}%
  \begin{subfigure}{.5\textwidth}
  \centering
    \includegraphics[width=0.9\linewidth]{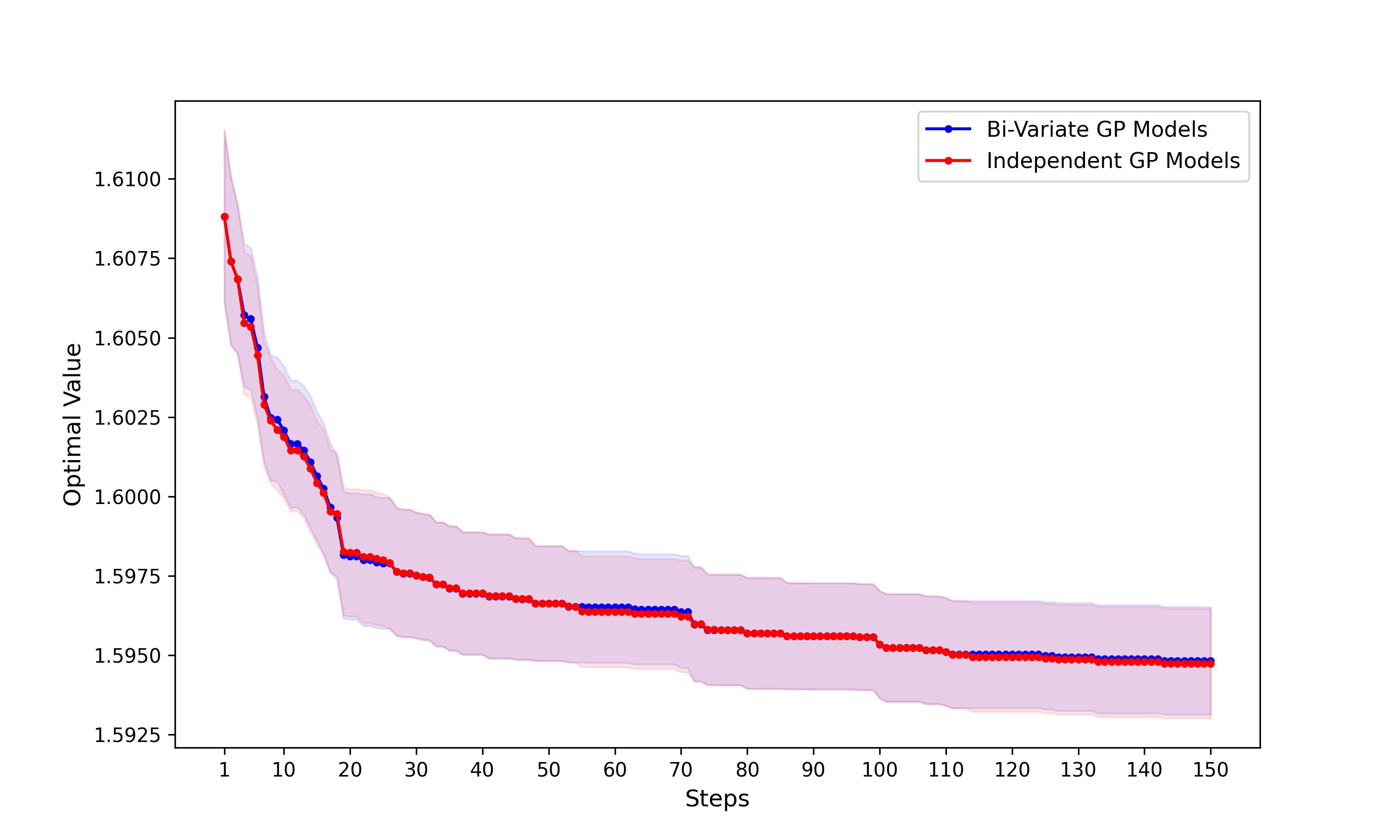}
    \caption{Initial Sample Size $50$.}
    \label{fig:50_initial_sample_size}
  \end{subfigure}
  \caption{Results from Different Initial Sample Sizes.}\
  \label{fig:case_study}
\end{figure}

We compare  the proposed Bayesian Optimization approach based on bivariate Gaussian process with constrained Bayesian Optimization developed by \cite{gardner2014bayesian}. In \cite{paper2}, constrained Bayesian Optimization has been applied to optimize the cure process, which outperforms some traditionally used derivative-free optimization approaches in \cite{limaye2024computational}. In the comparison of Bayesian Optimization approaches, micro-replication is often required to stabilize the random behavior of a single sample path. However, micro-replication of the real simulation can be numerically expensive. We use 882 simulation runs from the cure of the L-shaped laminate to fit polynomial functions to serve as the pseudo simulator (e.g., \cite{kerfonta2024sequential}). The pseudo simulator can facilitate micro-replication 
and provide robust comparison between Bayesian Optimization approaches. The two methods are described as follows:
\begin{itemize}
\item {\bf Independent Gaussian Process Model:} This is the approach developed in \cite{gardner2014bayesian}, and applied to cure process optimization in \cite{paper2}. The surrogate model is the independent Gaussian Process model in \eqref{eq:bivariate} and the acquisition function is given by \eqref{eq:EI_c}.
\item {\bf Bivariate Gaussian Process Model:} This is the proposed approach in this paper. The surrogate model is the bivariate Gaussian process in \eqref{eq:3covs} and the acquisition function is given by Proposition \ref{prop1}.
\end{itemize}
For both methods, we use an initial dataset to train the surrogate model, and then use the acquisition function to select new input point for 150 steps. The entire procedure is replicated for 30 times. In Figure \ref{fig:case_study}, we depict the mean and 95\% confidence band of the optimal value at each step for the cases with 25 and 50 initial data points, respectively.

From Figure \ref{fig:case_study}, we observe that the optimal values from both methods within 150 steps. Additionally, the converged optimal value decreases as the initial sample size increases. This indicates that larger initial samples provide more reliable starting points, leading to better final solutions. 
While both methods exhibit similar convergence behavior, the difference in final optimal values is minimal, suggesting comparable performance. However, the computation time for the proposed method is approximately three times longer than that of independent Gaussian Process, and the computational cost increases with larger initial sample sizes. The results shows that Bayesian Optimization with bivariate Gaussian process under the covariance function used in this paper may not necessarily benefit the target optimization problem. 

\section{Conclusion}
In this paper, we study the bivariate Gaussian process model to enhance constraint optimization by capturing dependencies between objective and constraints. In our numerical results,  
the proposed approach can not outperform  the method without characterizing the dependencies. To further improve the proposed approach, we can improve the covariance function by applying non-separable covariances \cite{gelfand2010handbook} function and utilizing the dependence structure to reduce the running time of the cure simulation.

\appendix 

\section*{Appendix: Proof of Proposition \ref{prop1}} 

We first provide a useful lemma on the expectation of truncated bivarite normal distribution. 
\begin{lemma}\label{lemma:bivariate_normal}
    Let  $(W, V)$  be a random vector following a bivariate normal distribution:
\begin{equation*}
(W, V) \sim \mathcal{N} \left( \begin{bmatrix} \mu_w \\ \mu_v \end{bmatrix}, \begin{bmatrix} \sigma_w^2 & \rho \sigma_w \sigma_v \\ \rho \sigma_w \sigma_v & \sigma_v^2 \end{bmatrix} \right),    
\end{equation*}
based on properties of truncated bivariate normal distribution \cite{rosenbaum1961moments}, we have that
\begin{align}
    \mathrm{E}[W \mid W \le w, V\geq v] =\mu_w+\rho \sigma_w \frac{\phi\left( \frac{v-\mu_v}{\sigma_v} \right)}{1-\Phi\left( \frac{c-\mu_v}{\sigma_v} \right)}-\sigma_w \sqrt{1-\rho^2} \frac{\phi\left( \frac{w-\mu_w-\rho \sigma_w \frac{\phi\left( \frac{v-\mu_v}{\sigma_v} \right)}{1-\Phi\left( \frac{v-\mu_v}{\sigma_v} \right)}}{\sigma_w \sqrt{1-\rho^2}} \right)}{\Phi\left( \frac{w-\mu_w-\rho \sigma_w \frac{\phi\left( \frac{v-\mu_v}{\sigma_v} \right)}{1-\Phi\left( \frac{v-\mu_v}{\sigma_v} \right)}}{\sigma_w \sqrt{1-\rho^2}} \right)}
    \label{eq:EYZ}
\end{align}
\end{lemma}
Following this lemma, we can obtain the closed-form expression of 
$\mathrm{E}(y(\bm x)|y(\bm x)\leq y_{min}, z(\bm x)\geq c)$ based on the bivariate normal distribution of $(y(\bm x), z(\bm x))$ for a fixed $\bm x$, and obtain the expression for ECI accordingly.



\end{document}